\documentclass[a4paper,10pt]{iopart}
\usepackage{color,graphicx,setspace,iopams}

\begin{document}

\title{Resonant tunnelling features in the transport spectroscopy of quantum dots}
\author{C C Escott$^{\dag}$, F A Zwanenburg and A Morello}
\address{Australian Research Council Centre of Excellence for Quantum Computer Technology, School of Electrical Engineering and Telecommunications, University of New South Wales, Sydney NSW 2052, Australia\\
$^{\dag}$ Present address: Sapphicon Semiconductor Pty Ltd.,
Homebush Bay NSW 2127, Australia}

\ead{a.morello@unsw.edu.au}

\begin{abstract}

We present a review of features due to resonant tunnelling in
transport spectroscopy experiments on quantum dots and single donors.
The review covers features attributable to intrinsic
properties of the dot as well as extrinsic effects, with a focus on
the most common operating conditions. We describe several phenomena
that can lead to apparently identical signatures in a bias
spectroscopy measurement, with the aim of providing experimental
methods to distinguish between their different physical origins. The
correct classification of the resonant tunnelling features is an
essential requirement to understand the details of the confining
potential or predict the performance of the dot for quantum information processing.
\end{abstract}

\section{Introduction}
With the rapid miniaturization of electronic devices comes the need
to understand their operation at an unprecedented level. Quantum
mechanical effects are becoming increasingly important in the design
of nanometre-scale transistors \cite{levi07}, posing new challenges
in the experimental characterization and corresponding theoretical
description of the devices that will constitute the next generations
of commercial electronics. Quantum effects in the electron transport
of nanodevices have already been the subject of many years of
fundamental research. Single-electron devices \cite{likharev99} are
the focus of much of this research, possessing highly non-linear
current-voltage characteristics resulting from a combination of
quantum mechanical (charge quantization, discrete energy levels) and
classical effects (Coulomb repulsion). These properties have lead to
single-electron devices being used as ultra-sensitive charge sensors
\cite{brenning06JAP} and amplifiers for quantum signals
\cite{devoret00N}. Here the term ``single-electron device'' refers
to the fact that the current consists of individual electrons,
sequentially flowing through a nanostructure.

Upon further miniaturization, one reaches the fully
quantum-mechanical regime of few-electron quantum dots
\cite{LeoFewEl} or single-donor devices
\cite{rogge,calvet07,khalafalla,tan10}. Both types of quantum wells
are characterized by an electron confinement strong enough to create
single-particle energy levels with a spacing larger than the thermal
broadening. This allows individual resolution of the energy levels
of the quantum dot, whether from electrostatically defined dots or
arising from a donor potential. In this work we will refer to both
types of devices as `(quantum) dots', unless specifically mentioned
otherwise. Few-electron quantum dots are the subject of intense
investigation in the context of quantum information processing
\cite{bennett00N}. Individual charge carriers confined in quantum
dot structures can be used to encode quantum information using
either the charge
\cite{fedichkin00NT,hayashi03PRL,hollenber04PRB-cq,petta04PRL,gorman05PRL}
or the spin
\cite{loss,kane,vrijen00PRA,friesen03PRB,elzerman04N,petta,cerletti05NT,hill05PRB,tokura06PRL,morello09PRB}
degree of freedom. For this purpose it is essential that the
electrons occupy well-defined energy states, and the details of the
excitation spectrum are vital in determining, for instance, the
coherence and relaxation times of the qubit
\cite{khaetskii00PRB,tahan05PRB,stano06PRB}.

\begin{figure}[h]
\centering
\includegraphics{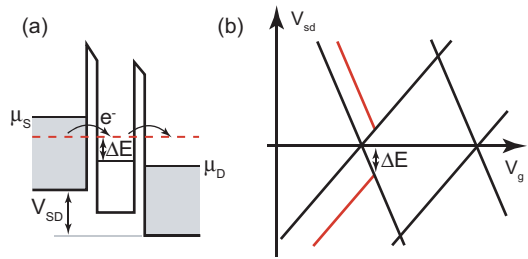}
\caption{(a) Schematic diagram of the electrochemical potential
levels of a dot coupled to source and drain reservoirs. (b) The bias
spectroscopy measurement consists of ground state transport lines
which determine the Coulomb diamond edges (black lines), and
resonant tunnelling features which appear as additional lines (red).
These are caused by other single-particle levels or resonant
processes at an energy $\Delta$E above or below the electrochemical
potential of the ground state, as indicated by the dashed line in
(a).} \label{fig:simple_tunnel} \noindent
\end{figure}

The most common experimental method to investigate the energy level
structure of a dot is bias spectroscopy \cite{sohn,hanson07}. For a
dot coupled to source and drain reservoirs, resonant tunnelling
occurs when occupied electronic states in the source and/or drain
align with available states in the dot, as sketched in Figure
\ref{fig:simple_tunnel}(a). This allows electrons to flow
sequentially through the dot and to be detected as a change in
source-drain current. Voltages on nearby gates are used to control
the electrochemical potential of the dot. A bias spectroscopy
experiment is a measurement of the source-drain conductance as a
function of source-drain bias and gate voltages. This yields a
2-dimensional map as shown in Figure \ref{fig:simple_tunnel}(b),
which contains diamond-shaped regions with well-defined charge
number on the dot (Coulomb diamonds). Resonant tunnelling features
induce a distinct change in current above normal device fluctuations
and appear as lines running parallel to the diamond edges. Such
features are commonly observed in single-electron devices, however
their explanation is often simplistic or lacking. This is a result
of there being many possible sources of resonant tunnelling
features, only identifiable via subtle differences in their
behaviour.

In this article we give an overview of the different resonant
tunnelling features expected in bias spectroscopy experiments of
single-dot devices. The discussion is restricted to the most common
operating conditions, corresponding to sequential single electron
tunnelling. We review methods for determining the nature of these
features with reference to experimental examples, and aim at
providing clear guidelines to distinguish between features of
different physical origin. We only consider shifts or extensions of
the lines, without discussing their width or shape. These methods
will serve as a resource for the engineer or physicist seeking to
explain measurement results or design next generation
single-electron devices.

\begin{figure}[h]
\centering
\includegraphics[width=0.8\textwidth]{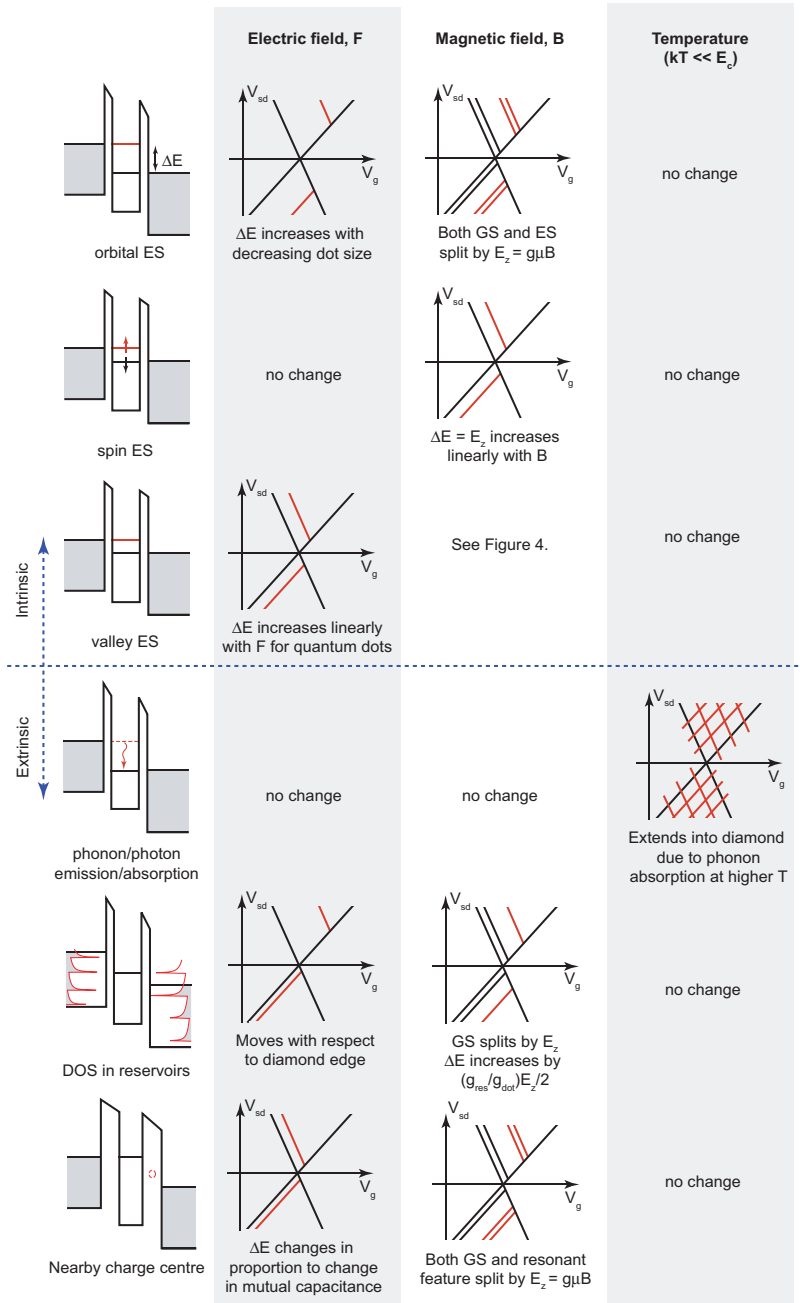}
\caption{Common sources of resonant tunnelling features in bias
spectroscopy experiments. Features may be due to intrinsic
properties of the dot [excited states (ES)], or due to extrinsic
effects. The response of each scenario to experimental variables is
summarized here to aid the identification of the source of resonant
tunnelling features. Explanations of each scenario and experimental
examples are given in the text.} \label{fig:summary} \noindent
\end{figure}

The paper is structured around Figure \ref{fig:summary}, which
summarizes the different conditions that lead to resonant tunnelling
features in bias spectroscopy, as well as methods to identify them.
The main distinction we draw is between ``intrinsic features'',
originating from the internal energy level structure of the dot, and
``extrinsic features'', appearing because of discrete levels or
quantized modes in the environment coupled to the dot.

\section{Intrinsic features}
\subsection{Orbital excited states}
Electrons are added to a dot when their potential is raised enough
to overcome the addition energy, the sum of the energy level spacing
and the Coulomb repulsion from electrons already on the dot
\cite{grabert92}. Detecting and mapping the addition of electrons is
precisely the purpose of a bias spectroscopy experiment, as
explained in the previous section. Excited states of the dot that
enter the bias window will increase the current due to the
introduction of an additional conductance channel. These states will
ideally appear as steps in the current, running parallel to the
Coulomb diamond edge. The change in current will not always be
step-like, since changing the source-drain bias or gate voltage
affects to some degree the tunnel barriers, the confining potential
and the coupling to reservoir states. Step-like behaviour will be
visible in only one biasing direction if the device contains
asymmetric tunnel barriers \cite{cobdenspin,bjork04}.

The excited state spectrum of a dot is a function of the material
and operating conditions (e.g. confinement strength and effective
mass), with energy level separation varying from the order of
$\rm{\mu}$eV for large quantum dots \cite{beenakker91} up to 100 meV
in ultrasmall dots \cite{zwanenburgultra}. In some simple cases, the
particle-in-a-box approximation \cite{sohn} can be used to calculate
the orbital level spacing. This simplification must however be
applied with caution, since it often produces a value that seems
consistent with experimental observations but could be explained
equally well by other phenomena.

Varying the size of the dot helps identify resonant tunnelling
features as being due to orbital excited states. Changing the dot
size will change its excited state spectrum. If the dot size is
defined electrostatically, for example through depleting a buried
charge layer, then changing the dot size is generally possible by
varying the nearby gate potentials.

In a simple rate equation model of sequential tunnelling through a
dot \cite{bonet02}, the current across the left barrier for positive
source-drain bias is given by,
\begin{eqnarray}
 I^{+} & = e\Gamma_{\textsc{l,g}}\rm{P_{\textsc g}}+\Gamma_{\textsc{l,e}}\rm{P_{\textsc e}}, \\
 & = e\frac{\Gamma_{\textsc{l,g}}\Gamma_{\textsc{l,e}}\left(g_{\textsc g}\Gamma_{\textsc{r,g}} + g_{\textsc e}\Gamma_{\textsc{r,e}}\right)}{\Gamma_{\textsc{l,g}}\Gamma_{\textsc{l,e}}+g_{\textsc g}\Gamma_{\textsc{r,g}}\Gamma_{\textsc{l,e}}+g_{\textsc e}\Gamma_{\textsc{r,e}}\Gamma_{\textsc{l,g}}},
\label{eq:rateeq}
\end{eqnarray}
where $\Gamma_{\textsc{r,g}}$ is the tunnel rate from the dot ground
state (\textsc{g}) through the right barrier (\textsc{r}), and
similarly for the excited state (\textsc{e}) and left barrier
(\textsc{l}). $\rm{P_{\textsc{e(g)}}}$ represents the probability of
occupying the excited (ground) state and $g_{\textsc{e(g)}}$ is its
degeneracy. If we assume that the tunnel rate into the ground state
is approximately equal to that for the excited state, then equation
(\ref{eq:rateeq}) simplifies to
\begin{equation}
 I^{+} = e\frac{\left(g_{\textsc g} + g_{\textsc e}\right)\Gamma_{\textsc r}\Gamma_{\textsc l}}{\Gamma_{\textsc l}+\left(g_{\textsc g} + g_{\textsc e}\right)\Gamma_{\textsc r}}.
\label{eq:possimple}
\end{equation}
Similarly the current under negative bias is given by
\begin{equation}
 I^{-} = e\frac{\left(g_{\textsc g} + g_{\textsc e}\right)\Gamma_{\textsc r}\Gamma_{\textsc l}}{\left(g_{\textsc g} + g_{\textsc e}\right)\Gamma_{\textsc l}+\Gamma_{\textsc r}}.
\label{eq:negsimple}
\end{equation}
Equations (\ref{eq:possimple}) and (\ref{eq:negsimple}) show that in
the presence of asymmetric tunnel barriers (e.g. $\Gamma_{\textsc
l}\gg\Gamma_{\textsc r}$), the magnitude of the current step
due to the addition of an excited state to the bias window will be
different for positive and negative source-drain bias. Conversely,
if the tunnel rate into the excited state is much higher than into
the ground state ($\Gamma_{\rm E}\gg\Gamma_{\rm G}$), then
asymmetric barriers will result in both positive and negative bias
currents $I^{-}$ and $I^{+}$ being dominated by the tunnel rate into
the excited state through the rate limiting barrier (i.e. if
$\Gamma_{\textsc l}\gg\Gamma_{\textsc r}$, then $I^{-} = I^{+} =
e\Gamma_{\rm R,E}$).

Higher order tunnelling events through dot states can provide
further information on the source of resonant tunnelling features
\cite{averin92}. Tunnelling processes simultaneously involving two
or more electrons (known as co-tunnelling) appear in bias
spectroscopy as lines traversing Coulomb diamonds where transport is
normally blocked. When co-tunnelling lines join resonant tunnelling
features in adjacent diamonds, those features must be due to an
intrinsic property of the dot. Co-tunnelling has been clearly
observed, for example, in a vertical few-electron InGaAs quantum dot
\cite{defranceschi01} as well as in lateral GaAs \cite{zumbhl04}
quantum dots.

The first experimental example of analysis of orbital excited states
of a quantum dot via bias spectroscopy was in 1992 \cite{johnson92}.
Here, gate voltage sweeps performed at different source-drain biases
showed signatures of 0D dot states, identified by their
correspondence with the calculated energy level spacing. A similar
study was completed soon after, generating the first full `stability
diagram' of conductance in the bias vs gate-voltage plane
\cite{foxman93}.

\begin{figure}[ht!]
\centering
\includegraphics[width=\textwidth]{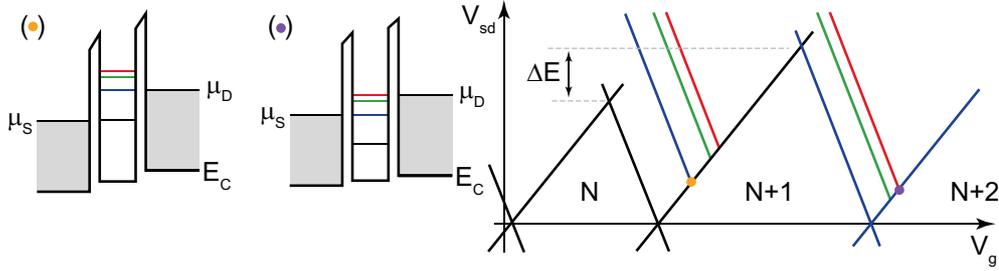}
\caption{Stability diagram for a dot with a constant charging energy
and a ladder of excited states (right panel). The orange spot
corresponds to the leftmost diagram of the electrochemical potential
levels. Here the ground state (black line) is aligned with the
source, while the first orbital excited state is aligned with the
drain (blue line). The same orbital forms the ground state at the
next electron transition (middle panel), where the first excited
state now corresponds to the green line. For clarity we show lines
of increased conductance in only one direction.}
\label{fig:orbital_ES_adj_diam} \noindent
\end{figure}

The excitation spectra and sizes of adjacent Coulomb diamonds can
provide additional evidence for orbital excited states. As mentioned
previously, the addition energy, i.e. the energy required to add one
electron to a dot, is equal to $E_{\textsc a} = E_C + \Delta E$,
where $E_C$ is the charging energy and $\Delta E$ the orbital
energy. The addition energies of two consecutive electrons entering
the same orbital are equal to $E_C + \Delta E$ and $E_C$. The
Coulomb diamond heights of the $N^{th}$ and $N+1^{th}$ diamond
should thus differ by the orbital energy, assuming a constant
charging energy. If higher orbital excited states appear, then the
spectra of the $N^{th}$ and $N+1^{th}$ diamond should be the same
except for the energy shift, as shown in Figure
\ref{fig:orbital_ES_adj_diam}. This effect has been observed very
clearly, for example, in carbon nanotubes \cite{cobdenshell} as well
as GaAs heterostructures \cite{stewart97} and Au nanoparticles
\cite{ralphau}.

\subsection{Spin excited states}
The behaviour of discrete quantum states in response to static
magnetic fields is well understood \cite{hanson07}. A magnetic field
breaks the spin degeneracy of the ground and excited states. The
resulting split states will be separated by the Zeeman enery $E_z =
g\mu_{\textsc b}B$, where $g$ is the electronic g-factor,
$\mu_{\textsc b}$ is the Bohr magneton and $B$ is the magnitude of
the applied magnetic field. For positive g-factor materials such as
silicon, the spin-down level becomes the ground state and spin-up
the excited state, separated by the Zeeman energy $E_{\rm z}$. The
net result is that the spin excited state moves at a rate $E_{\rm z}
(B)$ with respect to the Coulomb diamond edge, as shown in Figure
\ref{fig:summary}.

A clear example of spin excited states has been
reported in \cite{hansonzeeman}. The observed resonant tunnelling
features correspond to spin-split orbital excited states of a
single-electron GaAs quantum dot. Both the ground state and the
orbital excited state are spin-degenerate at 0 T, and split by the
Zeeman energy as a magnetic field is applied. Over the range from 6
T to 10 T the splitting of both ground and orbital states increases
by roughly the same amount.

Spin excited states have been studied comprehensively in GaAs dots
\cite{hanson07,Potok03,Kogan04,laurens}. Detailed measurements at
different magnetic fields show how the excited state moves with
respect to magnetic field according to $E_Z = g\mu_BB$. Spin excited
states of single dots have also been measured in metal nanoparticles
\cite{ralph95}, carbon nanotubes \cite{cobdenshell,jarillo}, InAs
nanowires \cite{bjork2}, Si nanowires \cite{zwanenburg}, Si/Ge
nanowires \cite{roddaro08} and donors in silicon
\cite{tan10,calvet08}.

\subsection{Valley excited states}
\label{sec:valley} In some materials degeneracies exist in the
carrier band, such as the six-fold conduction band degeneracy
in bulk silicon \cite{sze}. Restricting the momentum of electrons in
a silicon device by applying confining potentials or strain in
specific directions can lift this degeneracy
\cite{ando82,saraiva09,boykin04}. The energy separation of the
conduction band minima (valleys) then causes additional features in
a transport spectroscopy measurement, i.e. valley excited states. In
quantum dots, breaking all valley degeneracies results in a
complicated order of filling of electron states based on the relative size of the
orbital level spacing, interaction energies, Zeeman energy and
valley splitting. In addition, the valley-orbit interaction can lead
to a mixing of the orbital wave functions, such that orbitals and valleys
are no longer good quantum numbers \cite{friesen09}. The situation
is usually simpler in donor confinement potentials, where the orbital level spacings are
very large. For clarity, in the discussion below we shall assume the
existence of distinct orbital and valley quantum numbers.

\begin{figure}[ht!]
\centering
\includegraphics[width=\textwidth]{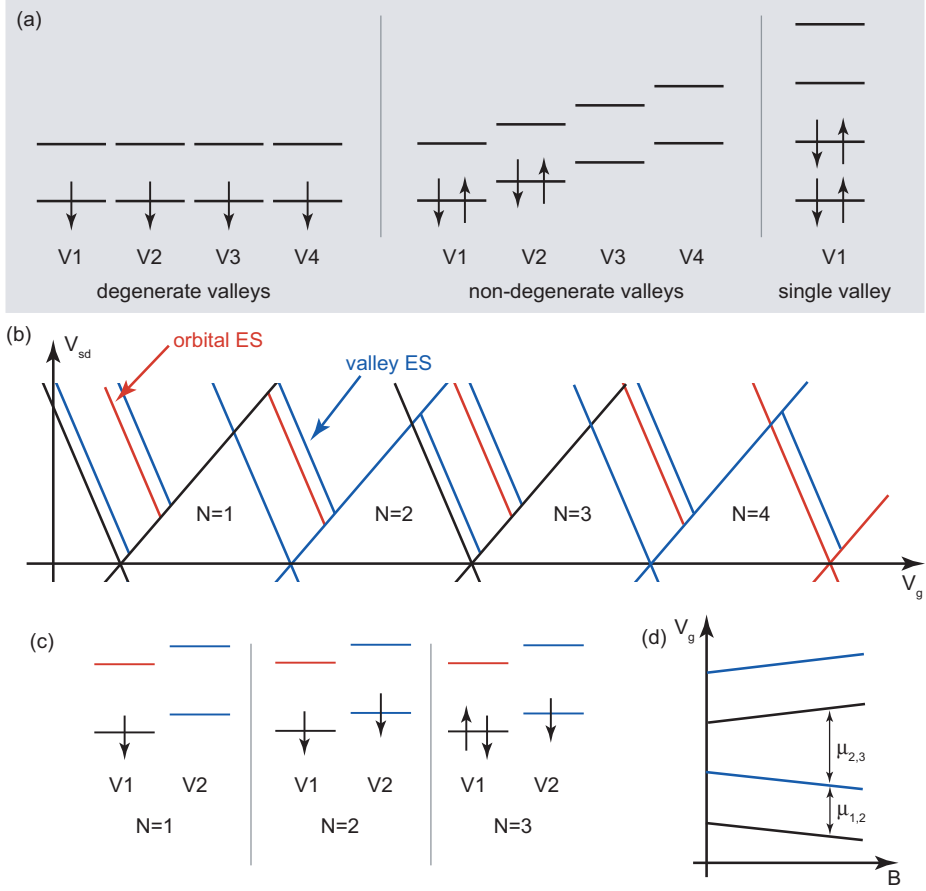}
\caption{(a) Possible spin configurations for 4 electrons in a
degenerate, four-valley (left), non-degenerate four-valley (centre),
and single valley (right) semiconductor. The exchange energy is
assumed smaller than the valley splitting and orbital level spacing.
(b) Coulomb diamonds showing the excited state spectrum expected for
the two, non-degenerate valley spin filling sequence shown in (c).
For clarity we show lines of increased conductance in only one
direction. Here, the exchange energy is considered larger than the
valley splitting so that consecutive electrons are added to
different valleys. (d) The electrochemical potentials, $\mu$, of the
spin filling shown in (c) as a function of magnetic field. Adding an
electron of opposite spin to the same orbital will increase the
Coulomb peak spacing for increasing magnetic fields, see e.g.
$\mu_{2,3} (B)$. } \label{fig:valleys} \noindent
\end{figure}

For degenerate multi-valley dots, electrons are added to the same
orbital in each of the valleys (Figure \ref{fig:valleys}(a)). Valley
excited states will be preferentially populated if the valley
degeneracy is broken and the exchange and orbital energies are
greater than the valley splitting, as in the sketch in Figure
\ref{fig:valleys}(c). This is due to exchange interaction being
negligible for electrons populating different valleys \cite{hada03}.
Here electrons are added to two different valleys consecutively,
resulting in a total spin S=1 for the N=2 state. This is in contrast
to the S=0 state that would result in a single valley dot, where the
N=2 ground state is a singlet (provided the orbital level spacing is
larger than the exchange energy \cite{stewart97,tarucha00}). The
magnetic field dependence of the zero bias Coulomb peaks can
therefore be used to help identify the origin of resonant tunnelling
features. As shown in Figure \ref{fig:valleys}(d), the
electrochemical potential $\rm{\mu}_{1,2}$ will not change with
increasing magnetic field when parallel spins are consecutively
added to a dot (here N=1$\to$2). Adding an electron of opposite spin
to the same orbital will however induce an increase in
electrochemical potential $\rm{\mu}_{2,3}$ with increasing magnetic
field. The spin filling of consecutive electrons follows from the
evolution of the Coulomb peak spacing in ground state
magnetospectroscopy \cite{folk}. This experimental method thus helps
to recognize whether the electrons fill different orbitals or
different valleys \cite{lim09}.

For a two-dimensional electron gas (2DEG) created in the [001] plane
in silicon, the confinement will be in the same direction as the
longitudinal effective mass for two of the six valleys. These two
z-valleys are therefore lower in energy than the four remaining
valleys. The degeneracy of the z-valleys may be broken for a
tightly-confined 2DEG, depending on the properties of the nearby
interface \cite{saraiva09,friesen09,takashina04,friesen07}. This
splitting can be as high as a few meV for strong electric fields and
enhanced by applying additional degrees of confinement
\cite{goswami07,sasaki09}. For common operating conditions, the
z-valley splitting is linear in the applied field
\cite{sham79,grosso96}.

Valley excited states are also shifted via an electric field due to
the Stark effect. For donors in silicon, electric fields exceeding
2-3 MV/m will lead to appreciable Stark shifting \cite{friesen05,
rahman09}. If the donor is near an interface, an accumulating
electric field can lead to the hybridization of the donor
wavefunctions with those of interface confined states
\cite{calderon06PRL,lansbergen08}, leading to vastly different
excited states from those observed in the bulk.

Splitting of the lowest two \textbf{k}$_z$ valleys in a silicon
inversion layer was first observed in 1966 \cite{fowler66}.
Subsequent experimental work in resonant tunnelling through quantum
wells has also led to observation of valley splitting in Si/Ge
\cite{monroe92,weitz96,koester97} and SiO$_2$/Si/SiO$_2$ devices
\cite{takashina04}. A comprehensive study of valley splitting in
silicon dots is lacking, although level filling of an SOI quantum
dot has been reported \cite{rokhinson01}. Here, the evolution of
Coulomb peaks with magnetic field reveals a filling of the
first five charges with alternating spin-down and spin-up electrons,
implying non-degenerate valleys as in the middle panel of
Figure \ref{fig:valleys}(a).

\section{Extrinsic features}
\subsection{Photon/phonon assisted tunnelling}

Photon and phonon emission and absorption are capable of enhancing
current through a dot by offering additional inelastic tunnelling
processes. A resonant tunnelling current upon emission of photons or
phonons relies on charge carriers tunnelling into or out of the dot
by an inelastic relaxation process. In the case of very low
temperatures, the number of populated phonon modes vanishes,
preventing the absorption of energy from the phonon bath. Upon
increasing the temperature, phonon modes become occupied and phonon
absorption allows current to flow through the dot when it would be
otherwise blockaded at $T=0$. Photonic/phononic excitations are most
easily identified as being independent of magnetic field.

\begin{figure}[ht]
\centering
\includegraphics[width=\textwidth]{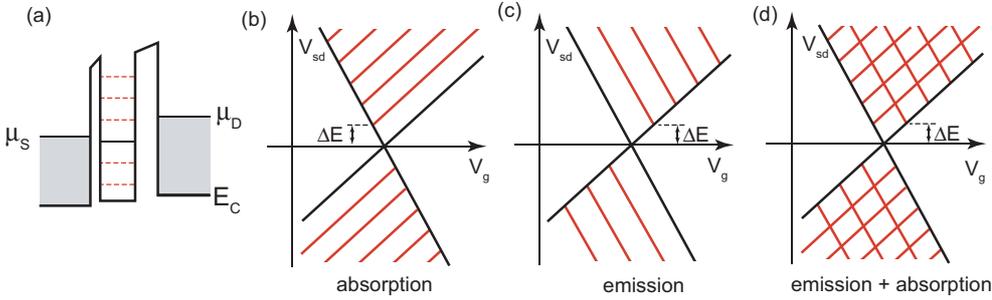}
\caption{(a) Schematic diagram of the electrochemical potential
levels of a dot with asymmetric tunnel barriers. Red dashed lines
indicate a discrete energy spectrum of phonons. In this case
resonant tunnelling features corresponding to (b) absorption or (c)
emission of phonons will appear parallel to only one diamond edge.
(d) For asymmetric tunnel barriers with \textsl{both} absorption and
emission processes available, features will appear parallel to both
diamond edges. The situation with symmetric barriers in the presence
of only absorption or emission will display a similar signature.}
\label{fig:phonon_asymm} \noindent
\end{figure}

If the process of relaxation or excitation possesses a discrete
energy spectrum that becomes resonant with a tunnelling event,
additional features will be observed in bias spectroscopy, see Figure
\ref{fig:phonon_asymm}. For the
case when a well-defined cavity exists within a device, a discrete
phonon energy spectrum will be present and allow enhanced electron
tunnelling through emission at discrete energies. The spacing of
resonant tunnelling features due to phonon emission is given by
\begin{equation}
 \Delta E_{\rm phon} = {\rm h}c/\lambda,
\end{equation}
where $c$ and $\lambda$ are respectively the speed and wavelength of the
photon/phonon. In the case of a closed (open) cavity the wavelength
$\lambda$ is equal to 2$L$ (4$L$), where $L$ is the length of the
photon/phonon cavity that the tunnelling electrons are coupled to.
The corresponding wavelengths thus follow from the energies $n\Delta
E$ at which the resonances are observed.

In case of phonon emission, many roughly equidistant resonant
tunnelling features appear at energies $n\Delta E$. An energy
emission of $n\Delta E$ can be described by two possible scenarios:
(i) one phonon with energy $n\Delta E$ is emitted. Each line
corresponds to one phonon mode of energy $n\Delta E$, hence the
modes are equidistant in energy as in a harmonic oscillator
potential. Such a regular set of phonon modes may well have similar
shapes, resulting in comparable electron-phonon couplings and
current peak heights for all modes. (ii) $n$ phonons with energy
$\Delta E$ are emitted. The latter process requires a very strong
electron-phonon coupling \cite{flensberg}.

For symmetric tunnel barriers, resonant tunnelling features due
solely to emission to a discrete phonon spectrum run parallel to
both Coulomb diamond edges. In the presence of asymmetric tunnel
barriers, lines will be visible in only one direction for absorption
(Figure \ref{fig:phonon_asymm}(b)) and emission (Figure
\ref{fig:phonon_asymm}(c)). If both phonon emission and absorption
are present, lines will appear in both directions as shown in Figure
\ref{fig:phonon_asymm}(d).

Raising the temperature to stimulate phonon absorption will give a
conclusive answer as to whether phonon-assisted tunnelling is the
source of resonant features. Conductance peaks in carbon nanotube
quantum dots extend into the Coulomb blockaded regions upon
increasing the device temperature \cite{leturcq}. Phonon-assisted
tunnelling has also been shown to increase the conductance both by
spontaneous emission, e.g. \cite{vandervaart95,fujisawa,weber10} and
absorption, e.g. \cite{naber}. In the latter experiment the absorbed
phonons were on-chip generated surface acoustic waves. Another
experimental example is the observation of phononic excitations in
Si nanowires \cite{zwanenburg}. Here, the absence of the magnetic
field dependence of the resonant tunnelling features provides strong
evidence that their origin was phonon emission. Others have explored
features in transport spectroscopy of suspended GaAs quantum dots
where low-bias current is suppressed \cite{weig04}. The `phonon
blockade' is due to the formation of a `dressed' dot-phonon state
resulting from strong electron-phonon coupling. Bias spectroscopy of
single molecule transistors has also allowed the observation of
excited states due to vibrational modes of the molecule
\cite{park00,yu04}. Detailed numerical calculations were able to
confirm that the energies of certain molecular vibrational modes
matched the observed feature spacing. Similarly, an excitation
spectrum of suspended carbon nanotubes was found to be consistent
with calculation of its vibrational modes, and an order of magnitude
smaller than the dot orbital levels \cite{sapmaz06}.

In a manner analogous to phonon absorption, electromagnetic
radiation (i.e. photons) causes a peak in current when an applied
field becomes resonant with an otherwise prohibited tunnelling
event, see \cite{vanderwiel02_photons} and references therein. This
photon-assisted tunnelling (PAT) appears in a bias spectroscopy
measurement as an extension of the orbital excited state line into
the Coulomb diamond, resulting in a similar stability diagram to
that shown for phonon absorption in Figure \ref{fig:summary}.
Observation of PAT through excited dot states is in fact usually
achieved at a single (often zero) bias
\cite{oosterkamp97,meyer07,prati09}. The results in
\cite{oosterkamp97} demonstrate PAT in a GaAs quantum dot, observing
a dot orbital excited state at zero bias in addition to satellite
peaks caused by absorption of the microwave photons. A more recent
experimental example of PAT through single dots is the pumping of
electrons through excited dot states in carbon nanotubes via
application of a high frequency field \cite{meyer07}. The frequency
independence of certain resonant tunnelling features whilst the
device was under irradiation was identified as a signature they were
in fact due to excited states of the dot and not due to the
reservoirs.  Photon-assisted resonant tunnelling through a device
consisting of a single donor coupled to source and drain leads has
also been demonstrated \cite{prati09}. Excited states of the donor
were observed at zero bias by irradiating the device with a 40 GHz
AC signal. Earlier work focused on the additional features
introduced by photon absorption in a large quantum dot, but was not
used to probe the dot's excited states
\cite{kouwenhoven94_PAT,kouwenhoven94_PAT2}.

\subsection{DOS of the reservoirs}
As discussed earlier, bias spectroscopy requires a source and drain
of charge carriers in order to probe the states of the dot. The
properties of these reservoirs will strongly influence the behaviour
of the device, and in some cases be themselves the origin of
resonant tunnelling features. The DOS of the reservoirs within the
bias window determines the availability of charge carriers for
tunnelling, hence singularities in the reservoir DOS will appear in
the bias spectroscopy.

For the case where the source and drain reservoirs are
electrostatically induced two-dimensional charge layers, e.g. from
modulation doped heterostructures or gate induced
inversion/accumulation layers, the confinement perpendicular to the
2DEG plane quantizes the $z$-momentum of the carriers into distinct
subbands. Discontinuous changes in the 2D DOS occur when a different
subband becomes occupied. Only the lowest subband is occupied for
high confining fields or low temperatures \cite{lundstrom06}. The
DOS in this 2DEG subband is independent of energy (i.e. constant)
\cite{sze}, and therefore the reservoirs do not contribute any sharp
transport feature. However, if the width $W$ of the reservoirs
becomes comparable to the mean free path, $\ell$, quasi-1D DOS
features arise. We use the term `quasi-1D' since, if many 1D
subbands are occupied, a 2D DOS is recovered. The elastic mean free
path is proportional to the mobility, $\mu$ of the 2DEG through,
\begin{equation}
 \ell = \nu_{\textsc f}\tau = \hbar k_{\textsc f}\mu/e,
\end{equation}
where $\nu_{\textsc f}=\hbar k_{\textsc f}/m^*$ is the Fermi
velocity, $k_{\textsc f}$ is the Fermi wave vector, $m^*$ is the
effective mass and $\tau=m^*\mu/e$ is the elastic scattering time.
Typical values for the mean free path are in the range $30-120~$nm
in silicon and $\sim10~\mu$m in GaAs 2DEGs. For reservoirs with
width comparable to or less than $\ell$, the number of states per
subband is given by,
\begin{equation}
 N(E) = \frac{1}{\pi\hbar}\sqrt{\frac{2m^*}{E-E_n}},
\label{eq:DOS}
\end{equation}
where $E_n$ is the subband energy. Equation (\ref{eq:DOS}) shows
that a singularity in the DOS is reached at energies equal to any of
the subbands. Experimentally, these DOS peaks will be asymmetric,
rising sharply then falling proportional to $1/\sqrt{E_{\textsc
f}}$, where $E_{\textsc f}$ is the Fermi energy. However scattering
off lattice, impurity and interface defects often broadens these
peaks and make it difficult to observe their $1/\sqrt{E_{\textsc
f}}$ rolloff.

The spacing of the subband energy levels can be approximated in a
number of ways. For an electron accumulation/inversion layer, the
subbands due to motion perpendicular to the interface are well
approximated by the eigenvalues of a triangular well
\cite{davies98},
\begin{equation}
 E_{n_z}-E_c = \left[ \frac{\left((3/2)\pi(n_z-1/4)\right)^2eF\hbar}{2m^*}\right]^{1/3}.
\end{equation}
Here $E_c$ is the conduction band minimum, $F$ is the confining
electric field and $n_z$ is the subband index. In silicon, the
longitudinal effective mass applies to electrons populating the
\textbf{k}$_{\rm z}$-valleys and the transverse effective mass to
electrons in the \textbf{k}$_{\rm x}$- and \textbf{k}$_{\rm y}$-
valleys. By changing the magnitude of the confining field two
effects take place: 1) the increase in confinement strength changes
the subband energies; and 2) the conduction/valence band edge moves
relative to the Fermi level. The latter effect will dominate, since
changing the field scales the $n_z$ indexed subbands by $F^{1/3}$
and has little effect on the width of the inversion layer.
Therefore, directly gating the reservoirs helps identifying resonant
tunnelling features originating from the source or drain DOS, since
they will move relative to the Coulomb diamond edge. This is
summarized schematically in Figure \ref{fig:summary} and has
recently been demonstrated experimentally \cite{mottonen09}. With
asymmetric tunnel barriers, only resonant tunnelling features
relevant to the rate limiting barrier will be observed, leading to
features running in only one direction in a bias spectroscopy
measurement.

\begin{figure}[ht]
\centering
\includegraphics[width=\textwidth]{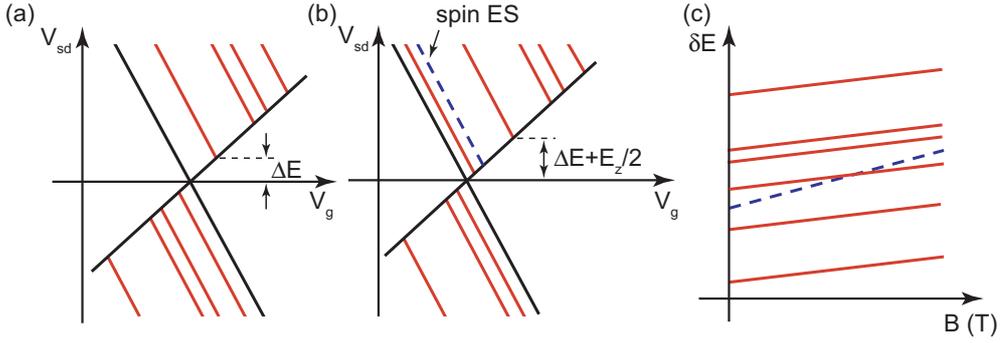}
\caption{Behaviour of resonant tunnelling features due to reservoir
DOS peaks (a) before and (b) after applying a magnetic field. (c)
Measuring the shift of features relative to the Coulomb diamond edge
as in \cite{mottonen09} provides a clear signature of reservoir DOS
peaks. For clarity we show lines of increased
conductance in only one direction. } \label{fig:DOS} \noindent
\end{figure}

In a magnetic field, the DOS of the reservoirs and the energy
spectrum of the dot split by the Zeeman energy. The splitting
will be equal if both reservoir and dot possess the same $g$-factor,
meaning features in the transport spectroscopy measurement due to
the reservoirs will not split. This is in contrast to spin excited
states of the dot itself, see Figure
\ref{fig:DOS}(a) and (b). Furthermore, the lower energy spin states
of the dot (including the Coulomb diamond edge) will shift down in
voltage by E$_{\rm z}$/2 while the DOS features will not move.
Plotting the shift of all the features relative to the Coulomb
diamond edge will show dot states splitting and/or moving by E$_{\rm
z}$ whereas features originating from the reservoir will move by
E$_{\rm z}$/2, as shown in Figure \ref{fig:DOS}(c).

Transport spectroscopy features have been attributed to structure in
the DOS for many years, for example the quasi-1D features observed
in narrow MOSFET channels
\cite{warren86,takeuchi91,matsuoka94,morimoto96} and carbon
nanotubes \cite{wilder98,schnenberger99,venema00}. The influence of
the reservoir DOS was first seen in the pioneering work on discrete
energy levels in metal nanoparticles \cite{ralph95}. Here, the BCS
DOS of the superconducting leads were visible as peaks in the
tunnelling current through a single Al particle. The first
experimental observations of features in resonant tunnelling through
semiconductor quantum dots attributable to peaks in the reservoir
DOS were achieved in 1997 \cite{schmidt97,kouwenhoven97}. At a later
stage others have made similar observations
\cite{bjork04,zumbhl04,lansbergen08,fuhrer}, although a study of
these features was not the primary objective of their work.

Fluctuations in the local DOS of disordered reservoirs attributed to
interference of diffusive charge carrier wavefunctions also cause
features in the transport spectroscopy of quantum dots. Since these
features are dominated by disorder, their behaviour in a magnetic
field is complicated \cite{koeneman01,pierre09} if not random
\cite{schmidt96}. They are also independent of devices size and
temperature \cite{holder00}.

\subsection{Nearby charge centres}
\begin{figure}[ht]
\centering
\includegraphics[width=\textwidth]{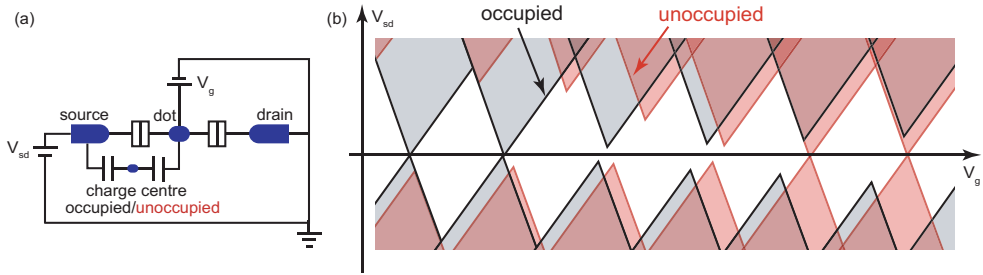}
\caption{(a) Circuit diagram of a quantum dot with source, drain and
gate. Additionally, a nearby, strongly coupled charge centre is
sketched as a parasitic dot carrying no current. The occupation of
the centre changes the electrochemical potential of the dot and
shifts its Coulomb diamonds, as sketched in (b). The overlap of the
Coulomb diamonds for an occupied and unoccupied charge centre result
in resonant tunnelling features in bias spectroscopy.}
\label{fig:charge_centre} \noindent
\end{figure}

The change in population of a charge centre located in the tunnel
barrier or near the reservoirs can also give rise to resonant
tunnelling features in bias spectroscopy measurements
\cite{hofheinz,pierre08}. These features are strongly dependent on
the capacitive coupling between the charge centre, the dot and the
gate, as well as the proximity of the charge centre energy levels to
the reservoir Fermi level. For weakly coupled traps, the resonant
features will be the same on all Coulomb diamonds \cite{pierre08}.
Strongly coupled charge centres however, produce a more obvious
signature where the change in charge centre occupancy results in
additional diamonds appearing alongside or in between the original
diagram, as shown in Figure \ref{fig:charge_centre}. In a magnetic
field, the states of both the dot and the charge centre will split
by the Zeeman energy, behaving in the same way as orbital states of
the dot. Characterization of the coupling capacitances enables
extraction of the location of the charge centre with respect to the
gates, and detuning of such charging events. Detailed experimental
studies of this type of resonant tunnelling event have been
undertaken \cite{hofheinz,pierre08,boehm05}. We note that the
coupling between a dot and a nearby charge centre has also been
proposed as a method to read out the spin of the charge centre
\cite{morello09PRB}.

\section{Conclusions}
Resonant tunnelling features in the bias spectroscopy of quantum
dots can arise from a broad range of physical phenomena. We have
shown that the response of a quantum dot to experimentally accessible
parameters, such as electric and magnetic fields or the temperature,
can be used to identify the nature of the tunnelling features. This
task is important for a variety of reasons, ranging from the basic
recognition of the nature of the binding potential due to dopants
\cite{lansbergen08}, to the assessment of the feasibility of quantum
computing schemes in quantum dots \cite{culcer09}. Therefore, the
information obtained from bias spectroscopy experiments can form the
first step towards understanding and predicting the properties of
quantum dots.

\ack This work was supported by the Australian Research Council, the
Australian Government, the US National Security Agency (NSA) and the
US Army Research Office (ARO) under contract number
W911NF-08-1-0527.

\end{document}